\documentclass[a4paper,12pt]{article}
\usepackage{amssymb,amsmath,epsfig}
\usepackage{hyperref}

\newcommand{\RR}{{\mathbb{R}}}

\newcommand{\pa}{\partial}

\newcommand{\dd}{{\rm d}}

\title{Helical buckling of Skyrme-Faddeev solitons}
\author{David Foster${}^\dagger$ and Derek Harland${}^\ast$
  \bigskip
  \\${}^\dagger$ \small{\textit{Instituto de F\'{i}sica de S\~{a}o Carlos, Universidade de S\~{a}o Paulo,}}
  \\ \small{\textit{Caixa Postal 369, CEP 13560-970, S\~{a}o Carlos-SP, Brazil}}
  \\ \small{\texttt{email address: dfoster@ursa.ifsc.usp.br}}
  \bigskip
  \\${}^\ast$ \small{\textit{Department of Mathematical Sciences, Loughborough University,}}
  \\ \small{\textit{Loughborough, Leics., LE11 3TU,  UK}}
  \\ \small{\texttt{email address: d.g.harland@lboro.ac.uk}}
  }
\date{17th February 2011}

\begin{document}

\maketitle

\abstract{Solitons in the Skyrme-Faddeev model on $\RR^2\times S^1$ are shown to undergo buckling transitions as the circumference of the $S^1$ is varied.  These results support a recent conjecture that solitons in this field theory are well-described by a much simpler model of elastic rods.}

\section{Introduction}
\label{sec:1}

It has recently been conjectured \cite{HSS2011} that solitons in a particular field theory, the Skyrme-Faddeev model \cite{Faddeev1975}, are well-described by an effective model based on Kirchhoff elastic rods.  It was shown in \cite{HSS2011} that the elastic rod model gives a good qualitative approximation and a reasonable quantitative approximation to low-charge Skyrme-Faddeev solitons.

The discovery of this elastic rod model motivates the search for elastic phenomena, such as buckling, in the Skyrme-Faddeev model.  In the present article we will investigate one such buckling effect in both the elastic rod and the Skyrme-Faddeev models, caused by the simultaneous stretching and twisting of a length of elastic rod (or soliton).  On a technical level, the most convenient way to stretch elastic rods and Hopf solitons is to place them on the manifold $\RR^2\times S^1$.  The rod is arranged to wind once around the $S^1$, and can be stretched by varying the circumference $P$ of $S^1$.

\begin{figure}[htb]
\begin{center}
\includegraphics[scale=1.0]{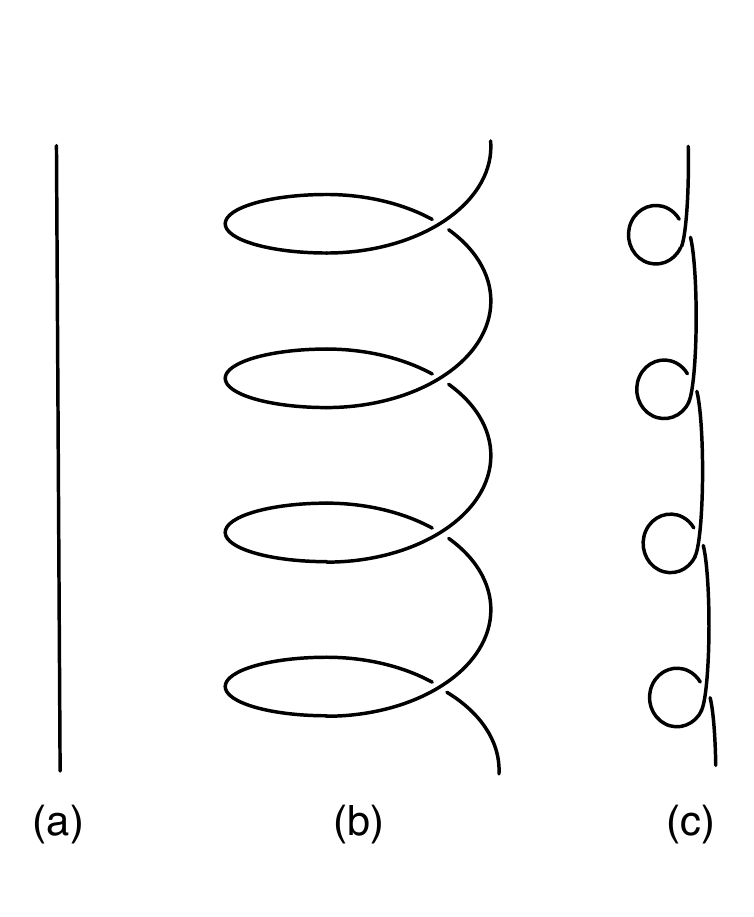}
\end{center}
\caption{Sketch of helical buckling}
\label{fig0}
\end{figure}

The kinds of effects that can occur are sketched in figure \ref{fig0}.  The simplest configuration consists of a straight rod (a).  If it is twisted in a suitable way, this straight rod may buckle at some critical value of $P$ to form a helix (b).  This helix could buckle again at another value of $P$, forming a kinked configuration (c).

The straight rod (a) is fixed by the group $\mbox{SO}(2)\times\mbox{SO}(2)$, where one copy of SO(2) acts on $S^1$ and the other by rotation on $\RR^2$.  The helix (b) is invariant only under a diagonal subgroup SO(2), while the kinked configuration has no continuous symmetries.  Thus the buckling transitions are examples of spontaneous symmetry breaking.

The motivation behind our research is twofold.  On the one hand, we were interested to further test the reliability and utility of the elastic rod model as a description of Skyrme-Faddeev solitons.  The rod model has already been successfully used to describe solitons on $\RR^3$, and $\RR^2\times S^1$ seems a good place to test it further.

On the other hand, we wanted to explore what seem to be fairly important aspects of the Skyrme-Faddeev model.  Solitons in this model resemble knotted rods, so it seems a good idea to investigate basic properties of these rods, such as their response to stretching and twisting.  Some studies of the Skyrme-Faddeev model on $\RR^2\times S^1$ have appeared before \cite{Hietarinta:2003vn,Jaykka:2009ry}, but, surprisingly, the simple buckling effects described here have not previously been investigated.  The Skyrme-Faddeev model has been proposed as a model of glueballs \cite{Faddeev:1998eq}.

An outline of the rest of this article is as follows.  In section \ref{sec:2} we will review the Skyrme-Faddeev and elastic rod models on $\RR^3$, and the connection between them.  In section \ref{sec:3} we explain our conventions for putting these models on $\RR^2\times S^1$, and discuss in detail the topological charges of Skyrme-Faddeev solitons on this space.  In section \ref{sec:4} we present some analytical and numerical methods that can be used to study the elastic rod model on $\RR^2\times S^1$.  In section \ref{sec:5} we present our main results, including direct comparisons between the two models.  We draw our conclusions in section \ref{sec:6}.

\section{The Skyrme-Faddeev model and elastic rods}
\label{sec:2}

\subsection{The Skyrme-Faddeev model}

The Skyrme-Faddeev model is an O(3) sigma model in 3+1 dimensions, whose Lagrangian is augmented by an additional term quartic in derivatives.  We will only be interested in static states, so that the fields of the model can be taken to be a triplet of scalars $\phi^a=(\phi^1,\phi^2,\phi^3)$, which are functions of $\vec x\in\RR^3$, and which satisfy the constraint $\phi^a\phi^a = 1$.  The static theory is defined by specifying the energy functional,
\begin{equation}
 E_{SF} = \frac{1}{32\pi^2\sqrt{2}} \int_{\RR^3} \Big( \pa_i\phi^a\pa^i\phi^a + \frac{1}{2} F_{ij}F^{ij} \Big) \dd^3 x,
\end{equation}
where
\begin{equation}
 F_{ij} = \epsilon_{abc}\phi^a\pa_i\phi^b\pa_j\phi^c.
\end{equation}
Static states in the model are solutions of the Euler-Lagrange equations associated to $E_{SF}$.

A configuration $\phi^a(\vec x)$ has finite energy only if $\pa_i\phi^a\to0$ as $|\vec x|\to\infty$.  Hence, for finite-energy configurations the limit of $\phi^a(\vec x)$ as $|\vec x|\to\infty$ is a well-defined point on the 2-sphere.  Therefore finite-energy configurations can be extended to continuous maps from $S^3=\RR^3\cup\{\infty\}$ to $S^2$.  It is well known that such maps have a topological invariant $Q\in\pi_3(S^2)\cong\mathbb{Z}$, known as the Hopf degree or topological charge.

The Hopf degree can be calculated in one of two ways.  The first possibility involves looking at the preimages of points on $S^2$ under the map $\phi^a$.  Generically these preimages will be unions of disjoint loops in $\RR^3$.  The Hopf degree $Q$ is equal to the linking number of the preimages of two distinct points \cite{BT1982}.  Alternatively, $Q$ may be calculated using an integral formula.  For any finite-energy configuration the tensor $F_{ij}$ defines a 2-form on the 3-sphere.  This is closed, and also exact since $H^2(S^3)=0$.  Therefore one can find a 1-form $A$ such that $F=\dd A$, and $Q$ is equal to the integral,
\begin{equation}
 Q = \frac{1}{16\pi^2}\int_{S^3} A\wedge F.
\end{equation}

The most important problem in the Skyrme-Faddeev model is the identification of stable static states.  Numerical simulations indicate that for each value of $Q$ there exists a unique energy-minimising configuration.  It is known that the energies of these configurations scale like $Q^{3/4}$; more precisely, it was proven in \cite{LY2004,VK1979} that there exist constants $C_1$ and $C_2$ such that
\begin{equation}
 C_1 Q^{3/4} \leq \inf_Q E_{SF} \leq C_2 Q^{3/4},
\end{equation}
where the infimum is taken over fields with Hopf degree $Q$.  Conjecturally, the value of the constant $C_1$ can be taken to be 1 \cite{Ward:1998pj}.

\subsection{Elastic rods}
\label{sec:2.2}

In \cite{HSS2011} it was shown that stable static states in the Skyrme-Faddeev model are well-approximated by elastic rods.  An elastic rod consists of two vector-valued functions $\vec x,\vec m$ of a real parameter $\sigma$.  The first function $\vec x$ specifies the location of the centreline of the rod in $\RR^3$.  The second function $\vec m$ specifies how the rod is twisted, and must satisfy the constraints,
\begin{equation}
 \vec m(\sigma)\cdot\vec m(\sigma) = 1, \quad \vec m(\sigma)\cdot \vec x'(\sigma) = 0.
\end{equation}
For small $\varepsilon$, the function $\vec x(\sigma)+\varepsilon\vec m(\sigma)$ describes the location in $\RR^3$ of a straight line in the material of the rod close to the centreline.

The energy functional for elastic rods is $E_R=AL+E_K$, where
\begin{equation}
 L = \int |\vec x'|\dd\sigma
\end{equation}
is the length of the rod and
\begin{equation}
 E_K = \int \left( B\kappa^2 + C\gamma^2 \right) |\vec x'|\dd\sigma
\end{equation}
is Kirchhoff's energy functional.  Here $\kappa$ is the curvature of the rod, defined by
\begin{equation}
 \kappa(\sigma) = \frac{|\vec t'|}{|\vec x'|},
\end{equation}
where
\begin{equation}
 \vec t(\sigma) = \frac{\vec x'}{|\vec x'|} 
\end{equation}
is the unit tangent vector to the rod.  On the other hand, $\gamma$ is the twist rate of the rod, and is defined by
\begin{equation}
 \gamma(\sigma) = \frac{\vec t\cdot \vec m' \times\vec m}{|\vec x'|}.
\end{equation}
By construction, the elastic rod energy functional is independent of the parametrisation.  It is often convenient to choose a parametrisation for which $|\vec x'|=1$, in which case the parameter $\sigma$ is denoted $s$ and called the arclength parameter.  The energy functional is invariant under SO(2) rotations of the normal vector $\vec m$, generated by
\begin{equation}
 \delta\vec m = \vec t\times\vec m.
\end{equation}

A slightly different way of describing elastic rods involves the Frenet frame.  The Frenet frame can be defined when the centreline $\vec x(s)$ is arclength-parametrised, with arclength parameter $s$.  It consists of three vectors $\vec t(s)$, $\vec n(s)$ and $\vec b(s)$, where $\vec t$ is the unit tangent vector, $\vec n=\vec t'/\kappa$, and $\vec b=\vec t\times\vec n$.  These vectors satisfy the Serret-Frenet equation:
\begin{equation}
 \frac{\dd}{\dd s} \left( \begin{array}{c} \vec t\\ \vec n\\ \vec b \end{array}\right)
 = \left( \begin{array}{ccc} 0 & \kappa & 0 \\ -\kappa & 0 & \tau \\ 0 & -\tau & 0 \end{array}\right)
 \left( \begin{array}{c} \vec t\\ \vec n\\ \vec b \end{array}\right),
\end{equation}
where $\tau(s)$ is a real function known as the torsion.  As long as $\kappa\neq0$ the material frame vector of a rod can be written $\vec m=\sin\alpha\, \vec n + \cos\alpha\,\vec b$ for some real function $\alpha(s)$.  Then the twist rate is
\begin{equation}
 \gamma(s) = \alpha'(s)-\tau(s).
\end{equation}

In order to obtain a better match with Hopf solitons, we will impose a non-intersection constraint on our elastic rods.  We will assume that the rods have a circular cross-section of radius $\rho$, and demand that the rods do not intersect themselves.  According to \cite{LSDR1999}, this is equivalent to the following two conditions:
\begin{enumerate}
 \item $\kappa(s)\leq \rho^{-1}$ for all $s$,
 \item $I\geq \rho$, where
 \begin{equation*}
  I = \frac{1}{2}\mbox{min}\left\{ d(s_1,s_2) | s_1\neq s_2 \mbox{ and }(s_1,s_2)\mbox{ is a critical point of }d\right\}
 \end{equation*}
 and $d(s_1,s_2)$ is the distance between the points $\vec x(s_1),\vec x(s_2)$.
\end{enumerate}
The first of these conditions essentially says that the radius of curvature $\kappa^{-1}$ of the centreline cannot be less than $\rho$.

\subsection{Elastic rods from Skyrme-Faddeev solitons}

The purpose of this article is to compare energy minima in the Skyrme-Faddeev and elastic rod models.  In order to make this comparison, we need to explain how configurations in the two models are related.  We do this by specifying a map from finite-energy field configurations in the Skyrme-Faddeev model to configurations of elastic rods.  A map sending elastic rod configurations to Skyrme-Faddeev field configurations was described in \cite{HSS2011}.

First of all, the centreline of the rod can be defined to be the preimage under $\phi^a$ of a point in $S^2$ antipodal to the asymptotic value of $\phi^a$.  More concretely, by making an SO(3) rotation of $\phi^a(\vec x)$ we can arrange that the following boundary condition is satisfied:
\begin{equation}
 \phi^a(\vec x) \to (0,0,1) \mbox{ as } r\to\infty.
\end{equation}
Then the centreline of the rod is defined to be the collection of all points $\vec x$ such that $\phi^a(\vec x)=(0,0,-1)$.  Typically this set will consist of a number of closed loops, each of which can be parametrised as $\vec x_I(\sigma_I)$ with $I$ an index labelling the loop.  Since the loops are closed, the parameter $\sigma_I$ can be chosen to lie in a closed interval $[\sigma^0_I,\sigma^1_I]$, in such a way that
\begin{equation}
 \vec x_I(\sigma^0_I) = \vec x_I(\sigma^1_I).
\end{equation}

The material frame vector $\vec m_I(\sigma_I)$ is obtained by projecting the vector
\begin{equation}
 \frac{\pa \phi^1}{\pa x^i}(\vec x_I(\sigma_I))
\end{equation}
orthogonally onto the space perpendicular to $\vec x_I'(\sigma_I)$, and normalising.  The obtained functions $\vec m_I(\sigma_I)$ satisfy
\begin{equation}
 \vec m_I(\sigma^0_I) = \vec m_I(\sigma^1_I).
\end{equation}
Notice that the SO(2) rotations of the material frame vector $\vec m$ correspond to the $\mbox{SO}(2)\subset\mbox{SO}(3)$ which fixes the asymptotic value of $\phi^a$.  We define the charge $Q$ of a collection of elastic rods to be the linking number of the collection of curves $\vec x_I$ with the collection of curves $\vec x_I+\varepsilon\vec m_I$, for small enough $\varepsilon$.  Then the map from Skyrme-Faddev field configurations to elastic rod configurations obviously preserves $Q$.

In \cite{HSS2011} it was shown that, for suitable choices of the parameters $A,B,C,\rho$, minimum-energy configurations in the elastic rod model with some fixed value of $Q$ look similar to minimum-energy configurations in the Skyrme-Faddeev model with the same value of $Q$.  The match was obtained by choosing the dimensionless parameter $C/B$ to be
\begin{equation}
 C/B = 0.85,
\end{equation}
and fixing the rod thickness $\rho$ to be
\begin{equation}
 \rho = \sqrt{\frac{B+C}{A}}.
\end{equation}
The remaining two parameters correspond to choices of units of length of energy.  We will choose these so that the energy minima in the elastic rod model with $1\leq Q\leq7$ have similar sizes and energies to the corresponding solitons in the Skyrme-Faddeev model, as in \cite{HSS2011}.  This leads to
\begin{equation}
\label{parameters}
 A = 0.0872, B = 0.0671, C = 0.0571.
\end{equation}
These parameters give the charge 1 soliton the correct size.  They overestimate its energy (by about 10\%), and underestimate the energies of solitons with higher charge.

\section{Solitons wrapping a circle}
\label{sec:3}

We will study Skyrme-Faddeev solitons not on $\RR^3$, but on the space $\RR^2\times S^1$.  The configuration space is by definition the set of $S^2$-valued functions $\phi^a(\vec x)$ satisfying
\begin{equation}
 \phi^a(x^1,x^2,x^3+P) = \phi^a(x^1,x^2,x^3),
\end{equation}
for some $P>0$.  Static stable states will be local minima of the energy per period,
\begin{equation}
\label{SF energy}
 E_{SF} = \frac{1}{32\pi^2\sqrt{2}} \int_{\RR^2\times S^1} \Big( \pa_i\phi^a\pa^i\phi^a + \frac{1}{2} F_{ij}F^{ij} \Big) \dd^3 x.
\end{equation}
Configurations with finite energy per period must satisfy the boundary conditions,
\begin{eqnarray}
\label{bc r}
 \pa_r\phi^a &\to& 0 \mbox{ as }r\to\infty, \\
\label{bc z}
 \pa_z\phi^a &\to& 0 \mbox{ as }r\to\infty, \\
\label{bc theta}
 \pa_\theta\phi^a &\to& 0 \mbox{ as }r\to\infty,
\end{eqnarray}
where $r,\theta$ are polar coordinates on $\RR^2$ and $z=x^3$ is the coordinate in the periodic direction.  These boundary conditions imply that $\phi^a$ has not one but two conserved topological charges, as we now explain.

We begin by considering the boundary condition \eqref{bc r}.  This implies that $\phi^a$ has a well-defined limit as $r\to\infty$.  The limiting function $\phi^a_\infty$ may depend on $\theta$ and $z$.

The boundary condition \eqref{bc z} implies that $\phi^a_\infty$ does not depend on $z$, but may still depend on $\theta$.  It follows that the map $\phi^a$ can be extended to the compactification of $\RR^2\times S^1$ obtained by adding a circle parametrised by $\theta$ at infinity.  This compactified space is in fact the 3-sphere.  One way to see this is to consider the following map from $\RR^2\times S^1$ to $S^3$:
\begin{equation}
 (r,\theta,z) \to \left( \begin{array}{c} \frac{r}{\sqrt{1+r^2}}\cos\theta \\  \frac{r}{\sqrt{1+r^2}}\sin\theta \\ \frac{1}{\sqrt{1+r^2}}\cos\left(\frac{2\pi z}{P}\right) \\ \frac{1}{\sqrt{1+r^2}}\sin\left(\frac{2\pi z}{P}\right) \end{array} \right).
\end{equation}
The reader may verify that this is an injection, and that the boundary of the image of $\RR^2\times S^1$ is a circle parametrised by $\theta$.  So, any field configuration $\phi^a$ satisfying \eqref{bc z} extends to a map from $S^3$ to $S^2$, and hence has a Hopf degree $Q\in\mathbb{Z}$.  The Hopf degree may be calculated as above, either by determining the intersection number of the preimages of two distinct points, or by integrating $A\wedge F$ over $\RR^2\times S^1$.

Now consider the boundary condition \eqref{bc theta}.  This condition implies that $\phi^a_\infty$ does not depend on $\theta$, but does not rule out $z$-dependence.  It follows that $\phi^a$ can be extended to the compactification of $\RR^2\times S^1$ obtained by adding a $z$-dependent circle.  This compactified space is $S^2\times S^1$, since the 1-point compactification of $\RR^2$ is $S^2$.  Maps from $S^2\times S^1$ to $S^2$ have two topological charges \cite{Pontrjagin1941,AK2005}.  First, there is a charge $D\in\mathbb{Z}$, which is equal to the degree of a map from $S^2$ to $S^2$ obtained by restricting to a slice of constant $z$ (this is independent of the choice of slice).  And second, there is a Hopf degree $Q'\in\mathbb{Z}_D$.

Finite-energy configurations satisfy all three conditions \eqref{bc r}, \eqref{bc z} and \eqref{bc theta}, so have three charges $Q,Q',D$.  These three charges are not independent: $Q'$ is equal to the value of $Q$ modulo $D$.  Therefore there are two independent topological charges, $Q,D\in\mathbb{Z}$.

In the present article we will restrict attention to configurations with $D=1$.  The corresponding elastic rods will have only 1 strand and will consist of functions $\vec x,\vec m$ of $\sigma \in\RR$, satisfying
\begin{equation}
\label{rod bc}
 \vec x(\sigma+\sigma_P) = \vec x(\sigma) + (0,0,P),\quad \vec m(\sigma+\sigma_P) = \vec m(\sigma),
\end{equation}
for some $\sigma_P>0$.  The energy per unit period is
\begin{equation}
 E_R = \int_0^{\sigma_P} \left( A + B\kappa^2 + C\gamma^2 \right) |\vec x'|\dd\sigma.
\end{equation}
Here we will investigate $Q=1,2,3$; Skyrme-Faddeev solitons with $D=1$ and $5\leq Q\leq8$ have previously been studied in \cite{Hietarinta:2003vn}.  Skyrme-Faddeev solitons with $D>1$ have been investigated in \cite{Jaykka:2009ry}, and these correspond to elastic rods with $D$ strands.

\section{Buckling of elastic rods}
\label{sec:4}

In this section we will outline numerical and analytical methods that can be used to study the buckling of elastic rods on $\RR^2\times S^1$.  We begin by considering in subsection \ref{sec:4.1} a straight rod winding around the $S^1$.  In subsection \ref{sec:4.2} we will consider a more general helical configuration, and derive conditions which determine when the straight rod may buckle to form helix.  In subsection \ref{sec:4.3} we briefly consider how the non-intersection constraints may be applied to the helix.  In subsection \ref{sec:4.4} we study buckling of the helix itself, and in subsection \ref{sec:4.5} we describe some numerical methods for studying elastic rods.

\subsection{Straight rod}
\label{sec:4.1}

We begin by making an ansatz:
\begin{equation}
\label{straight rod}
\begin{aligned}
 \vec x(s) &= (0,0,s) \\
 \vec m(s) &= (\cos(2\pi sQ/P), -\sin(2\pi sQ/P), 0).
\end{aligned}
\end{equation}
This ansatz satisfies the boundary conditions \eqref{rod bc} for elastic rods.  The Hopf degree is $Q$ and the second topological charge is $D=1$.  The ansatz has an $\mbox{SO}(2)\times\mbox{SO}(2)$ symmetry.  The first copy of SO(2) acts by translation on $z$ and rotates $\vec m$, and the second copy acts by rotation on $x,y$ and also on $\vec m$.  In fact, the ansatz is the unique ansatz with these symmetries, so by the principle of symmetric criticality it solves the equations of motion for elastic rods.  The energy per period of the ansatz \eqref{straight rod} is
\begin{equation}
\label{straight rod energy}
 E_R = AP + \frac{4\pi^2Q^2C}{P}.
\end{equation}

\subsection{First buckling}
\label{sec:4.2}

Now we consider a more general ansatz, which describes a helix with $M$ coils per period $P$:
\begin{equation}
\label{helix}
\begin{aligned}
 \vec x(s) &= \left( R\sin\frac{2\pi Ms}{L}, R\cos\frac{2\pi Ms}{L}, \frac{Ps}{L} \right) \\
 \vec m(s) &= \sin\alpha(s)\,\vec n(s) + \cos\alpha(s)\,\vec b(s) \\
 \alpha(s) &= \frac{2\pi (Q-M)s}{L}.
\end{aligned}
\end{equation}
Here $R>0$ is a parameter describing the radius of the coils of the helix, and $L$ is the length of the rod, given by the formula
\begin{equation}
 L^2 = P^2 + (2\pi MR)^2.
\end{equation}
The reader may verify that $|\vec x'|=1$, and hence that the rod is arclength-parametrised.

The Hopf degree of the configuration \eqref{helix} is independent of $R$, and is most easily evaluated in the limit $R\to 0$.  First of all, the vectors $\vec n$, $\vec b$ can be calculated to be
\begin{equation}
\begin{aligned}
\vec n(s) &= \left( -\sin\frac{2\pi Ms}{L}, -\cos\frac{2\pi Ms}{L}, 0 \right) \\
\vec b(s) &= \left( \frac{P}{L}\cos\frac{2\pi Ms}{L}, -\frac{P}{L}\sin\frac{2\pi Ms}{L}, -\frac{2\pi MR}{L} \right).
\end{aligned}
\end{equation}
In the limit $R\to 0$ these expressions reduce to
\begin{equation}
\begin{aligned}
\vec n(s) &= \left( -\sin\frac{2\pi Ms}{P}, -\cos\frac{2\pi Ms}{P}, 0 \right) \\
\vec b(s) &= \left( \cos\frac{2\pi Ms}{P}, -\sin\frac{2\pi Ms}{P}, 0 \right).
\end{aligned}
\end{equation}
Therefore in the limit $R\to0$ the ansatz \eqref{helix} reduces to
\begin{equation}\begin{aligned}
 \vec x(s) &= (0,0,s) \\
 \vec m(s) &= \sin\alpha\,\vec n + \cos\alpha\,\vec b = \left(\cos\frac{2\pi sQ}{P}, -\sin\frac{2\pi sQ}{P}, 0\right),
\end{aligned}
\end{equation}
which coincides with the ansatz \eqref{straight rod} for a straight rod.  In this limit the Hopf degree is obviously $Q$.

For $R>0$, the symmetry group of the ansatz \eqref{helix} is an SO(2) subgroup of $\mbox{SO}(2)\times\mbox{SO}(2)$.  When $R=0$ \eqref{helix} reduces to \eqref{straight rod}.  So the helix can be regarded as a symmetry-breaking perturbation of the straight rod.

It is convenient to introduce a dimensionless parameter $\lambda=P/L\in (0,1]$.  The radius and length can be recovered from $\lambda$ using the formulae $L=P/\lambda$, $2\pi MR=P\sqrt{\lambda^{-2}-1}$.  The energy per period of the helix is
\begin{equation}
\label{helix energy}
 E_R(\lambda) = \frac{AP}{\lambda} + \frac{(2\pi M)^2B}{P}\lambda(1-\lambda^2) + \frac{(2\pi M)^2C}{P}\lambda\left(\frac{Q}{M}+\lambda-1\right)^2.
\end{equation}
When $\lambda=1$ this reduces to the energy (\ref{straight rod energy}) of the straight rod.  We know that the straight rod is a critical point of the energy functional, but is it stable to small helical perturbations?

To answer this question, we just need to look at the slope of the function $E_R(\lambda)$ near $\lambda=1$.  We have
\begin{equation}
 \left. \frac{\dd E_R}{\dd\lambda}\right|_{\lambda=1} = - AP -\frac{8\pi^2M^2B}{P} + \frac{4\pi^2C}{P}(Q^2+2QM).
\end{equation}
The straight rod is unstable to the perturbation specified by $M$ if and only if this derivative is positive, or equivalently,
\begin{equation}
 A\left(\frac{P}{2\pi}\right)^2 < C(Q^2+2QM) - 2BM^2.
\end{equation}
The right hand side of this inequality does not depend on $P$.  If the right hand side is positive one can always find non-zero values of $P$ which satisfy the inequality, but if the right hand side is negative or zero the inequality can never be satisfied.  Therefore this inequality can be satisfied if and only if the right hand side is positive.  The right hand side is positive if and only if
\begin{equation}
 \frac{QC}{2B}\left(1-\sqrt{1+\frac{2B}{C}}\right) < M < \frac{QC}{2B}\left(1 + \sqrt{1+\frac{2B}{C}}\right).
\end{equation}
This means that for fixed $Q$ only finitely many values of $M$ give rise to instabilities of the straight rod.  For example, when $C/B=0.85$ and $Q=1,2,3$, only the following values of $M$ need to be considered:
\begin{equation}
\label{buckling}
\begin{array}{l|l}
 Q & M \\ \hline 1 & 1 \\ 2 & 1,2 \\ 3 & -1,1,2,3
\end{array}
\end{equation}
The value $P_c$ of $P$ at which the straight rod becomes unstable to a helical perturbation is given by
\begin{equation}
\label{critical period}
 P_c = 2\pi \sqrt{\frac{CQ^2+2CQM-2BM^2}{A}}.
\end{equation}

The energy \eqref{helix energy} blows up as $\lambda\to0$, therefore if the straight rod at $\lambda=1$ is unstable, the energy \eqref{straight rod energy} must attain a minimum at some value $\lambda_0\in(0,1)$.  Since the ansatz \eqref{helix} is fixed by symmetries, this minimum of $E_R(\lambda)$ corresponds to a solution of the equations of motion for the elastic rod model.

\subsection{Non-intersection constraint}
\label{sec:4.3}

As discussed in section \ref{sec:2.2}, demanding that an elastic rod does not self-intersect imposes two conditions on the configuration space of elastic rods.  For helical rods \eqref{helix}, the first condition $\kappa\leq \rho^{-1}$ is equivalent to
\begin{equation}
 \lambda^2(1-\lambda^2) \leq \left(\frac{P}{2\pi M\rho}\right)^2 .
\end{equation}
The left hand side of this inequality is bounded above by $1/4$, so if $P/M\geq\pi\rho$ this constraint does not restrict the range of $\lambda$.  If on the other hand $P/M<\pi\rho$ the constraint means that the allowed range of $\lambda$ is divided into two disjoint pieces $(0,\lambda_-]\cup[\lambda_+,1]$, where $0<\lambda_-<\lambda_+<1$ are defined by $2\lambda_\pm^2=1\pm\sqrt{1-(P/M\pi\rho)^2}$.

To understand the implications of the second constraint $I\geq\rho$, we first evaluate the distance function:
\begin{eqnarray}
 d^2(0,s) &=& \left| \left( R\cos\frac{2\pi Ms}{L}, R\sin\frac{2\pi Ms}{L}, \frac{Ps}{L} \right) - (1,0,0) \right|^2 \\
 &=& \left(\frac{P}{M}\right)^2 D\left(\frac{Ms}{L}\right) \\
 D(\sigma) &:=& \frac{1}{2\pi^2}\left(\frac{1}{\lambda^2}-1\right)\left(1-\cos2\pi\sigma\right) + \sigma^2 .
\end{eqnarray}
The function $D(\sigma)$ may or may not have a local minimum $\sigma_c\in[0,1]$, depending on the value of $\lambda$.  If a local minimum $\sigma_c$ exists, the second constraint is satisfied if and only if
\begin{equation}
 D(\sigma_c) \geq \left( \frac{2M\rho}{P} \right)^2
\end{equation}
If on the other hand $D$ has no local minimum, the constraint is satisfied.  For values of $\lambda$ close to 1, $D$ does not have a critical point, so a helix with $\lambda$ close to 1 always satisfies the second constraint.  For values of $\lambda$ close to 0, $D$ does have a local minimum $\sigma_c$.

When it exists, the value of $D(\sigma_c)$ is less than 1, and tends to 1 as $\lambda\to0$.  This means that if $P/M\leq 2\rho$, helices with $\lambda\approx0$ are completely ruled out (although helices with $\lambda\approx 1$ are still allowed).  This result matches geometric intuition: if $P/M$ is less than twice the thickness of the rod, a helix with large radius $R$ cannot avoid the overlapping of neighbouring coils.

In practise, the first constraint is more important than the second.  For large values of $P/M$ the energy function favours configurations with $\lambda\approx1$, so neither the first nor the second constraint influences the shape of the energy-minimiser.  If $P/M\leq\pi\rho$ the first constraint may influence the shape of the energy-minimiser: in particular, there will be two local energy minima, since the first constraint divides the range of $\lambda$ into two pieces.  The second constraint begins to affect the shape of the rod at even smaller values of $P/M$.

\subsection{Second buckling}
\label{sec:4.4}

We have described above how the straight rod, with $\mbox{SO}(2)\times\mbox{SO}(2)$ symmetry, can buckle to form a helix, with only SO(2) symmetry.  In this subsection we will describe how the helix can buckle again to form a kinked configuration with completely broken symmetry.  We will present two tools with which this second buckling can be studied: first, we will present an analytical method for calculating the critical period $P_c$ at which the buckling occurs; and second, we will describe some numerical methods with which the kinked configuration can be studied.

The analytical approach to the buckling is based on the treatment of Mitchell's instability of circular rods presented in \cite{Goriely2006}.  It can be shown that the Euler-Lagrange equations for the elastic rod energy functional are
\begin{equation}
\begin{aligned}
0 =& -A + B\kappa^2 + C \gamma^2 + 2B\left(\frac{\kappa''}{\kappa}-\tau^2\right) - 2C\gamma\tau \\
0 =& B(2\kappa'\tau+\kappa\tau') + C\gamma\kappa' \\
0 =& \gamma'.
\end{aligned}
\end{equation}
Here a prime denotes differentiation with respect to the arclength parameter $s$.  The helix \eqref{helix}, with energy $E_R(\lambda)$ given in \eqref{helix energy}, is a solution if and only if $\lambda$ solves the equation $\dd E_R/\dd \lambda=0$.  We will assume that $\lambda$ has been so chosen, and will ignore the restrictions imposed by the non-intersection constraints.  The curvature, torsion and twist rate for the helix are
\begin{equation}
\begin{aligned}
\kappa_0=& \frac{2\pi M}{L}\sqrt{1-\lambda^2} \\
\tau_0 =& -\frac{2\pi M}{L}\lambda \\
\gamma_0 =&\frac{2\pi}{L}(Q+M(\lambda-1)).
\end{aligned}
\end{equation}

Now we suppose that a small perturbation of the helix has been made, so that $\kappa=\kappa_0+\delta\kappa$, $\tau=\tau_0+\delta\tau$, $\gamma=\gamma_0+\delta\gamma$.  We will assume for simplicity that
\begin{equation}
\label{assumption}
\int_0^L \delta\kappa\,\dd s = 
\int_0^L \delta\tau\,\dd s = 0.
\end{equation}
The linearised equations of motion are equivalent to
\begin{equation}
0 = \left( \frac{\dd^2}{\dd s^2} + \kappa_0^2 + \left(\frac{C\gamma_0}{B}+2\tau_0\right)^2\right)\delta\kappa,
\end{equation}
with $\delta\tau=(C\gamma_0/B-2\tau)\delta\kappa/\kappa$ and $\delta\gamma=0$.  The linearised equations of motion have a solution $\delta\kappa\propto\sin(2\pi ns/L)$ only if there exists an integer $n$ such that
\begin{equation}
\label{buckling mode}
n^2 = M^2(1-\lambda^2) + \left( \frac{C}{B}(Q+M\lambda-M) -2M\lambda\right)^2.
\end{equation}
The existence of a solution to the linearised equations of motion indicates the presence of a buckling instability.  Thus it is straightforward to determine when buckling occurs: for each value of $P$ one computes the value $\lambda(P)$ of $\lambda$ which minimises the helix energy \eqref{helix energy}, and from this, the right hand side of \eqref{buckling mode}.  Buckling can occur at any value $P_c$ of $P$ for which the right hand side of \eqref{buckling mode} is the square of an integer.

The above discussion was based on the assumption \eqref{assumption}.  It can be shown with a little more work that dropping this assumption does not yield any additional solutions to the linearised equations of motion, essentially because perturbations for which $\delta\kappa,\delta\tau$ are constant correspond to modifications of the parameters $P,\lambda$ in the ansatz \eqref{helix}.

\subsection{Numerical methods}
\label{sec:4.5}

After the helix has buckled, it is no longer possible to obtain analytic solutions for elastic rods.  Instead, we employ numerical methods.  A discretisation of the elastic rods was presented in \cite{BWRAG2008}, which we briefly recall here.

The centreline of the rod $\vec{x}(\sigma)$ is replaced by a sequence of points $\vec x^i$ with $i\in\mathbb{Z}_N$.  The vector connecting two points is $\vec e_j:=\vec x^{j+1}-\vec x^j$, and the length of this vector is $l_j=|\vec e_j|$.  The material frame $\vec m(\sigma)$ is replaced by a sequence of vectors $\vec m_j$ of unit length satisfying $\vec m_j\cdot e_j=0$.

A unit tangent vector is defined by $\vec t_j := \vec e_j/l_j$.  We also introduce
\begin{equation}
\vec\Omega^j := \frac{2\vec e_{j-1}\times\vec e_j}{l_{j-1}l_j + \vec e_{j-1}\cdot\vec e_j}.
\end{equation}
The vectors $\vec\Omega^j$ approximate $\kappa(\sigma)\vec b(\sigma)$.  The discretised curvature is calculated from
\begin{equation}
(\kappa^j)^2 = |\vec\Omega^j|^2.
\end{equation}
Finally, a discretisation of the twist rate $\gamma(s)$ is given by $\gamma^i$, defined via
\begin{multline}
\sin \frac{2\gamma^j}{l_j+l_{j-1}} = \\(\kappa^j)^{-2} \left[ (\vec m_{j-1} \cdot \vec t_{j-1}\times\vec\Omega^j)(\vec\Omega^j\cdot \vec m_j) - (\vec m_{j} \cdot \vec t_{j}\times\vec\Omega^j)(\vec\Omega^j\cdot \vec m_{j-1}) \right].
\end{multline}

The discretisation of the elastic rod energy functional is now obviously
\begin{equation}
E_R = \sum_{j=0}^{N-1} Al_j + (B(\kappa^j)^2 + C (\theta^j)^2)\frac{l_j+l_{j-1}}{2}.
\end{equation}
It can be shown, with some effort, that this reduces to the usual energy functional in the continuum limit.  We searched for minima of this discretised energy using a simulated annealing algorithm.  The non-intersection constraint was imposed using the obvious discretisations of conditions 1 and 2 described in section \ref{sec:2.2}.  It was necessary to enforce the arclength parametrisation condition $l_i\approx l_j \forall i,j$, in order to maintain a good approximation throughout the simulation.  This was achieved by adding a penalty function to the energy.

\section{Skyrme-Faddeev solitons}
\label{sec:5}

In this section we will apply the methods of the previous section to study in detail elastic rods on $\mathbb{R}^2\times S^1$ with Hopf degree $Q=1,2,3$.  We then compare these predictions with full numerical simulations of the Skyrme-Faddeev model.

Our simulations of the Skyrme-Faddeev model were performed by evolving the models equation of motion on a discrete lattice of $300 \times 300 \times P/0.08$ points, with lattice spacing $\Delta x =0.08$. By varying the number of lattice points and the lattice spacing this size lattice was found to give the lowest energy solutions for all charges and periods $P$. The spacial derivatives were calculated on the lattice to fourth order. A Lagrangian multiplier was also included to preserve the constraint $\phi^a\phi^a = 1$.

\subsection{A straight Skyrme-Faddeev soliton}

Before discussing full numerical simulations, we first consider a straight Hopf soliton analogous to the straight rod \eqref{straight rod}.  Consider the following ansatz:
\begin{multline}
\label{straight ansatz}
(\phi^1,\phi^2,\phi^3) = \\ \left( \sin f(r) \cos\left(\theta + \frac{2\pi Qz}{P}\right), \sin f(r) \sin\left(\theta + \frac{2\pi Qz}{P}\right), \cos f(r) \right).
\end{multline}
This ansatz is invariant under a certain action $\mbox{SO}(2)\times\mbox{SO}(2)$, and is the most general ansatz with this symmetry.  Substituting this ansatz into the energy functional \eqref{SF energy} gives
\begin{equation}
\label{straight SF energy}
 E_{SF} = \frac{P}{16\pi\sqrt{2}}\int_0^\infty\left( (f')^2 + \left(\frac{1}{r^2} + \frac{4\pi^2 Q^2}{P^2} \right)\left(1 + (f')^2\right)\sin^2f\right) r\dd r.
\end{equation}
We impose the boundary conditions,
\begin{equation}
 f(0) = \pi,\quad f(r)\to0\mbox{ as }r\to\infty,
\end{equation}
so that the ansatz is well-defined at $r=0$ and can have finite energy per period.  Then the topological charge per unit period is $Q$.

For any values of $Q,P$, the minimum of \eqref{straight SF energy} with respect to variations in $f$ can be determined numerically, using a gradient flow algorithm.  In fact, it is sufficient to do this for $Q=1$, since the energy density in \eqref{straight SF energy} only depends on the ratio $Q/P$.

\subsection{Charge 1}

\begin{figure}[htb]
 \begin{center}
 \includegraphics[scale=1.0]{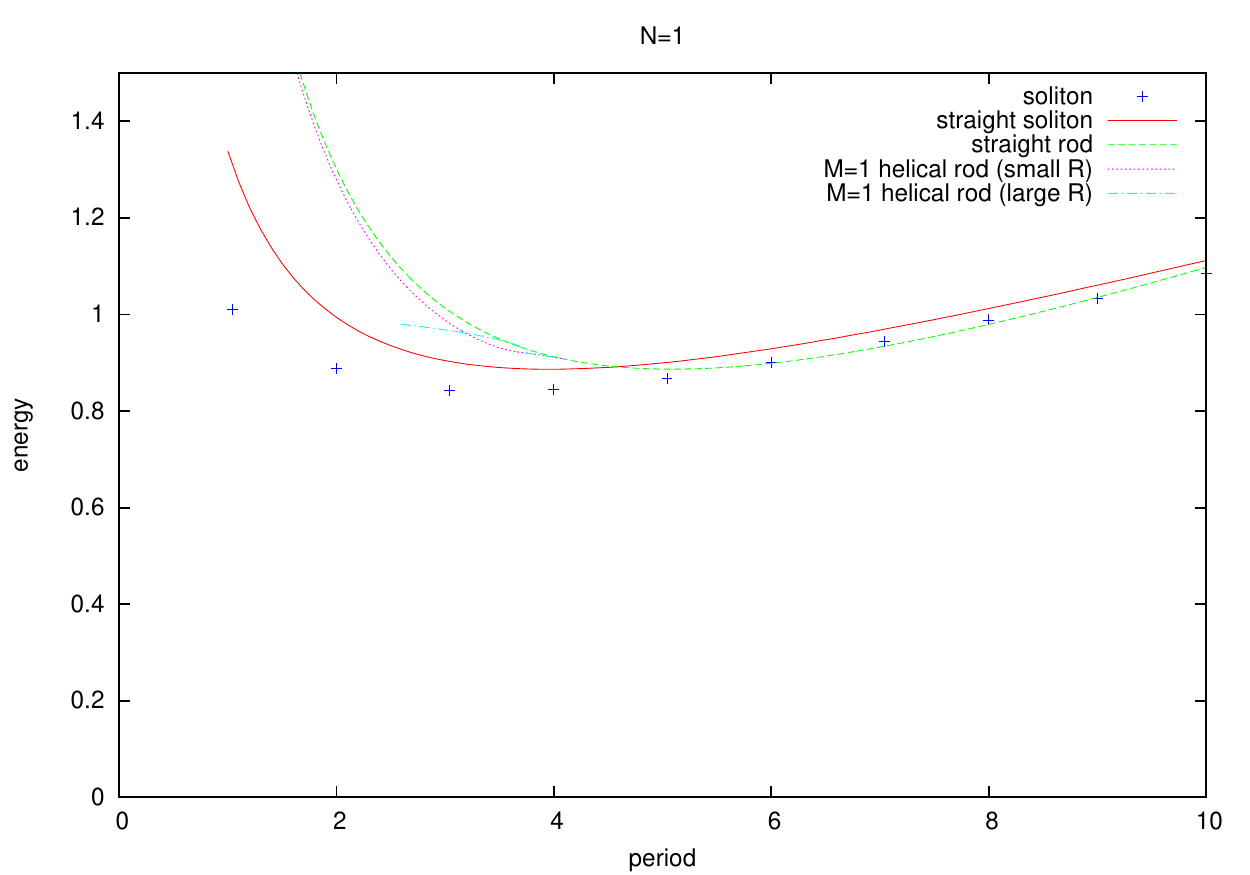}
 \end{center}
 \caption{Energies of solitons and rods with $Q=1$}
\label{fig1}
\end{figure}

\begin{figure}[htb]
\begin{center}
\includegraphics[height=2in]{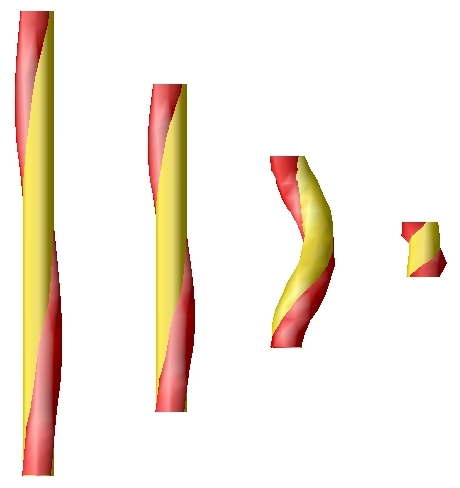}
\end{center}
\caption{Elastic rods with $Q=1$ and $P=7,5,3,1$.}
\label{fig2}
\end{figure}

\begin{figure}[htb]
\begin{center}
\includegraphics[height=2in]{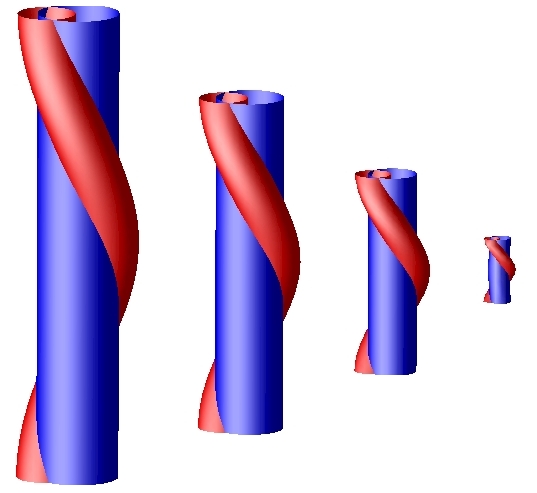}
\end{center}
 \caption{Solitons with $Q=1$ and $P=7,5,3,1$.}
\label{fig3}
\end{figure}

In figure \ref{fig1}, we have plotted the energies of elastic rods with Hopf degree $Q=1$ as a function of the period $P$.  For large enough periods, the only state in the elastic rod model is a straight rod \eqref{straight rod}.  As the period $P$ decreases, the energy of this state decreases, attains a minimum at $P=5.08$, and begins to increase.  At $P=4.09$ the straight rod becomes unstable and buckles to form a helix \eqref{helix} with $M=1$.  

At $P=3.75$, just after the helix has formed, the non-intersection constraint begins to play a role.  The space of allowable $\lambda$ is split into two intervals, and accordingly the elastic rod energy has two local minima, one being a helix with large radius $R$ and the other being a helix with small radius.  For most periods the helix with small $R$ has the lowest energy.  For a small range of periods $[2.59,3.19]$ the helix with large $R$ has the lowest energy.  However, when $P=2.59$ is reached the helix with large $R$ is no longer permitted by the non-intersection constraint, and once again the small-$R$ helix is the favoured configuration.

The main qualitative predictions of the elastic rod model are that for large periods the favoured configurations is a straight soliton, and for small periods the favoured configuration is a soliton in the shape of a helix with small $R$.  These states have been depicted in figure \ref{fig2}, with a yellow tube representing the curve $\vec x(s)$ and a red tube representing the curve $\vec x(s)+\varepsilon\vec m(s)$ for small $\varepsilon$.  There may also exist a helical soliton with large $R$ for a small range of periods, but since the corresponding state is short-lived in the rod model, one cannot be confident that it would exist in the Skyrme-Faddeev model.

In figure \ref{fig3} we have shown pictures of the Skyrme-Faddeev solitons with $Q=1$.  In these pictures, the blue surface represents the preimage under the map $\phi^a:\RR^3\to S^2$ of a circle surrounding the south pole $\phi^3=-1$ in $S^2$, and the red surface represents the preimage of a circle surrounding a point near the south pole.  The red tube links once with the blue tube, confirming that the charge is 1.  The solitons appear at first sight to be straight, however for smaller periods a slight buckling can be detected.  The pictures look similar to to the rods depicted in figure \ref{fig2}.  The main difference occurs at $P=3$, where the buckling of the rod is more pronounced than that of the soliton.

The numerically-determined energies of these solitons have been plotted in figure \ref{fig1}, as have the energies of straight solitons determined from \eqref{straight SF energy}.  It can be seen that the energies of Skyrme-Faddeev solitons determined using full numerical simulations are lower than those determined using \eqref{straight SF energy}.  For larger periods, the difference is small and can be attributed to numerical error (the energies determined from \eqref{straight SF energy} are more accurate than those determined using full numerical simulations).  For smaller periods, the difference is more pronounced, and occurs because the numerically-determined solutions are slightly buckled.

The elastic rod model gets the broad shape of the energy curve right, and energies are predicted with fairly good accuracy for periods in the range $[5,10]$.  The energy match for smaller periods is not so good, however.  This is not surprising, because the Skyrme-Faddeev model has many more degrees of freedom than the elastic rod model.  When an elastic rod is compressed beyond $P\approx5$, its energy increases, because the winding density $\gamma$ becomes large.  On the other hand, when a soliton is compressed beyond $P\approx5$, its energy stays roughly constant.  The soliton is able to maintain a low energy by changing its profile function $f(r)$: figure \ref{fig3} shows in particular that the soliton becomes very narrow at small periods.  Elastic rods do not have a degree of freedom analogous to the profile function, and this explains why rods approximate solitons poorly at low periods.  We conclude that elastic rods model $Q=1$ solitons well qualitatively, but the quantitative match is good only for a range of periods.

\subsection{Charge 2}

\begin{figure}[htb]
 \begin{center}
 \includegraphics[scale=1.0]{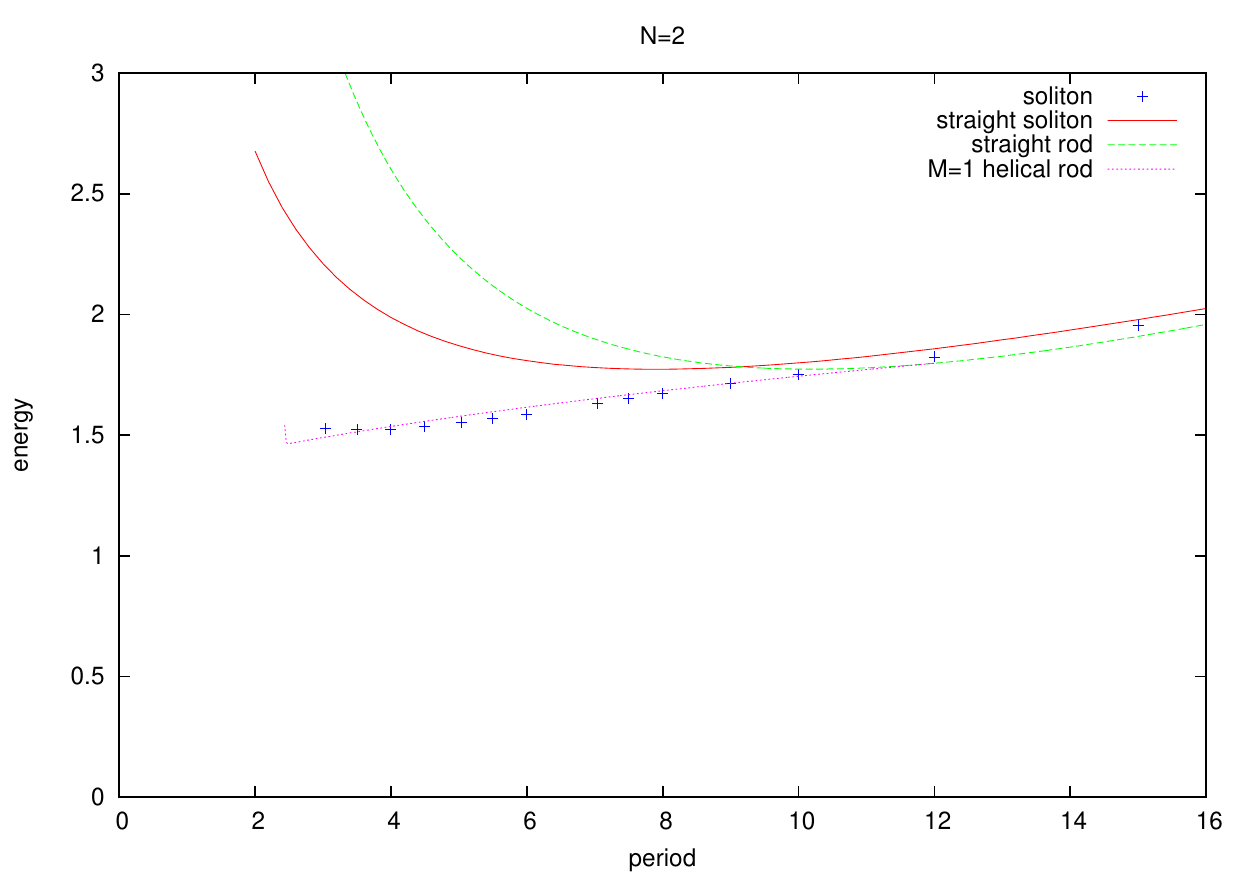}
 \end{center}
 \caption{Energies of solitons and rods with $Q=2$}
\label{fig4}
\end{figure}

\begin{figure}[htb]
\begin{center}
\includegraphics[height=3.5in]{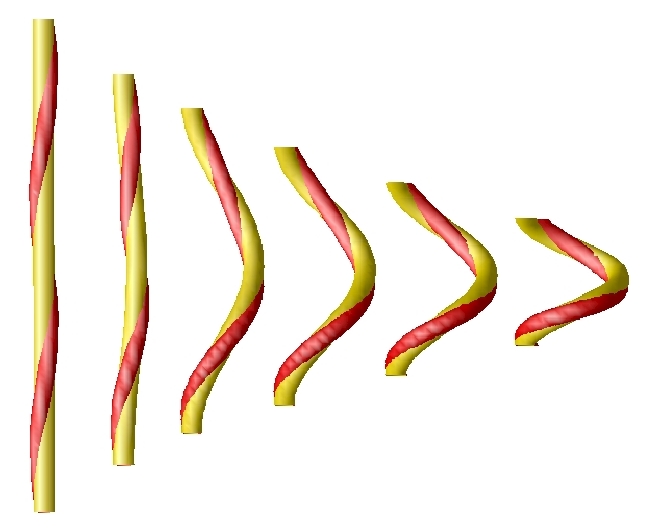}
\end{center}
\label{fig5}
\caption{Elastic rods with $Q=2$ and $P=15,12,10,8,6,4$.}
\end{figure}

\begin{figure}[htb]
\begin{center}
\includegraphics[height=3in]{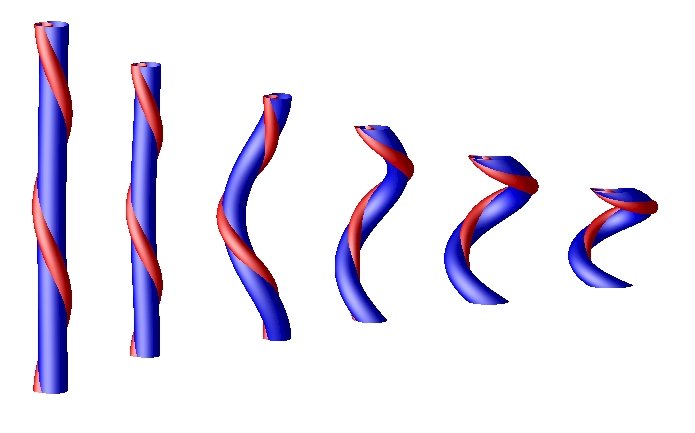}
\end{center}
 \caption{Solitons with $Q=2$ and $P=15,12,10,8,6,4$.}
\label{fig6}
\end{figure}

In figure \ref{fig4}, we have plotted the energies of elastic rods with Hopf degree $Q=2$ as a function of the period $P$.  For large enough periods, the only state in the elastic rod model is a straight rod \eqref{straight rod}.  As the period decreases the energy decreases, and at the critical period $P=12.0$ the straight rod buckles to form a helix with $M=1$.  The energy of the helix continues to decrease with the period until $P=2.49$ is reached, at which point the non-intersection constraint influences the shape of the rod and its energy begins to increase.  Table \eqref{buckling} indicates that there exists a helix with $M=2$, but this always has a higher energy than the $M=1$ helix and has not been plotted.  A sample of energy-minimising elastic rods have been depicted in figure \ref{fig5}.

Figure \ref{fig6} displays pictures of energy-minimising Skyrme-Faddeev solitons for a range of periods.  There is an excellent match with the elastic rods displayed in figure \ref{fig5}, including a buckling transition from a straight soliton to a helix at a period $P\approx12$.

The energies of the Skyrme-Faddeev solitons have been plotted in figure \ref{fig4}, along with the energy of a straight soliton determined from \eqref{straight SF energy}.  There is a small discrepancy between the energies determined from \eqref{straight SF energy} and those determined from full numerical simulations, and this can be attributed to numerical error.  For small periods the energies of the numerically-determined solitons are significantly lower than those of the straight solitons, reflecting the fact that the solitons in the numerical simulations are buckled.

It is clear from figure \ref{fig4} that the energies of Skyrme-Faddeev solitons and elastic rods with $Q=2$ are in remarkably good agreement.  We conclude that, for $Q=2$, elastic rods are a good model of Skyrme-Faddeev solitons, both qualitatively and quantitatively.

\subsection{Charge 3}

\begin{figure}[htb]
 \begin{center}
 \includegraphics[scale=1.0]{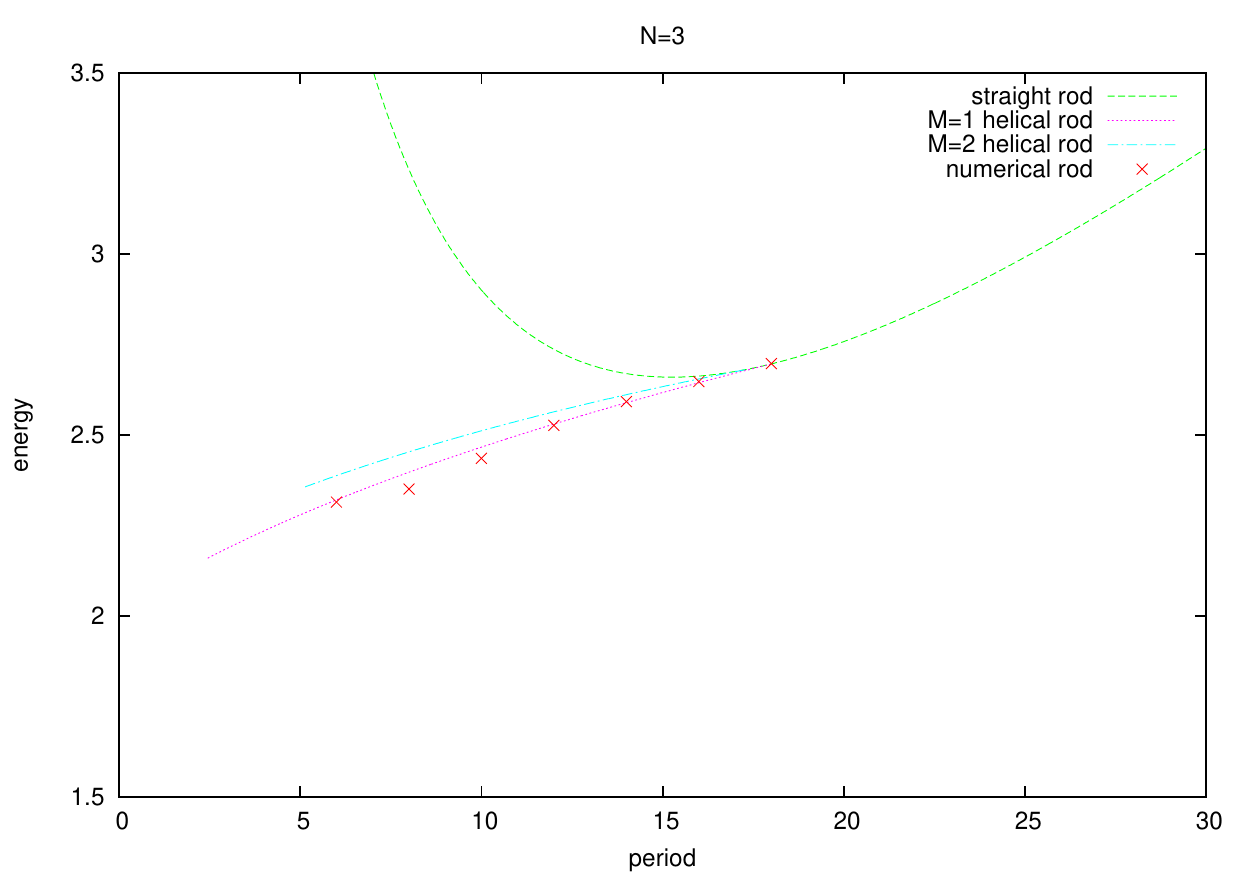}
 \end{center}
\caption{Energies of elastic rods with $Q=3$ as a function of $P$}
\label{fig7}
\end{figure}

\begin{figure}[htb]
 \begin{center}
 \includegraphics[scale=0.8]{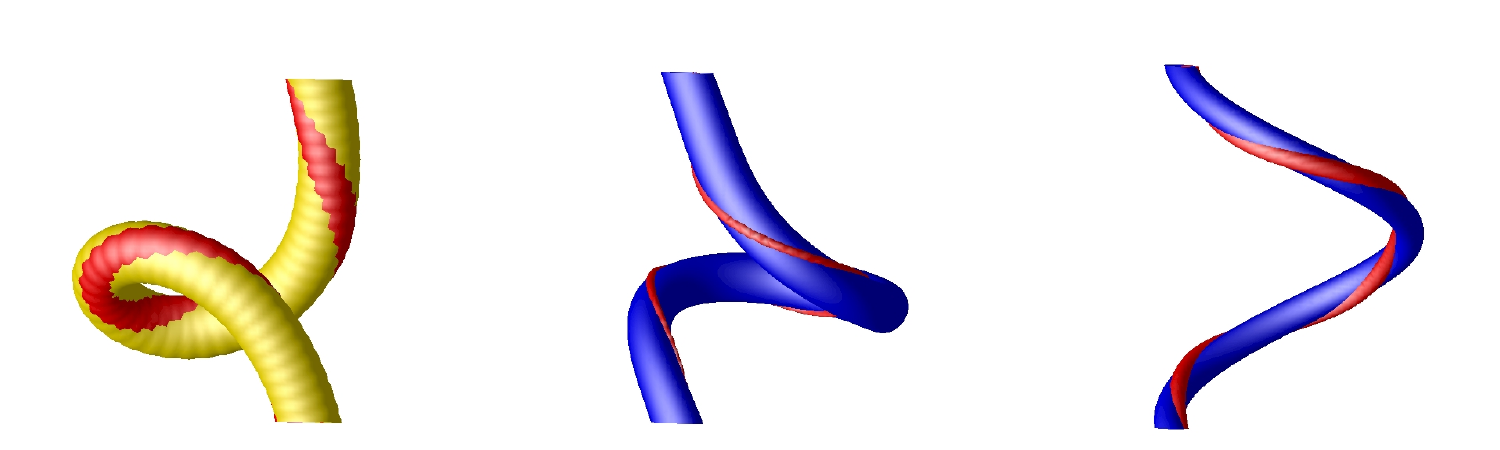}
 \end{center}
\caption{Solitons and rods with $Q=3$ and $P=6$.  The elastic rod is on the left, the kinked soliton solution is in the middle and the helical soliton solution is on the right.}
\label{fig8}
\end{figure}

In figure \ref{fig5} we have plotted the energies of elastic rods with Hopf degree $Q=3$ as a function of the period $P$.  For large enough periods, the only state in the elastic rod model is a straight rod \eqref{straight rod}.  As the period decreases the energy decreases, and at the critical period $P=18.03$ the straight rod buckles to form a helix with $M=1$.  The energy of the helix continues to decrease with the period, and at the critical period $P=13.46$ the helix buckles again to form a kinked configuration, as described in subsection \ref{sec:4.4}.  This kinked configuration cannot be constructed analytically, and has instead been constructed using the numerical methods described in subsection \ref{sec:4.5}.  The energies of the numerically-obtained rods dip below the helix energy when $P\leq13.46$, indicating that buckling has occurred.  We have also plotted the $M=2$ helical state, whose energy is slightly greater than that of the $M=1$ helix and the kinked configuration.  From \eqref{buckling} we see that there exist in addition helices with $M=-1,3$ but their energies are much greater than the other states and have not been plotted.

The elastic rod model predicts, therefore, that there should exist kinked Skyrme-Faddeev solitons with $Q=3$.  In order to test this prediction, we have simulated Skyrme-Faddeev solitons with $Q=3$ and $P=6$.  In fact, we found both a kinked soliton and a helical soliton.  Both of these are depicted in figure \ref{fig8}, along with the kinked elastic rod.  The kinked soliton has energy $E_{SF} = 2.285$ and the helix has energy $E_{SF}=2.282$.  Unfortunately the energies of the two solutions are within numerical accuracy, so we cannot conclude which has the lowest energy.  The helix was obtained by starting with a helical initial condition.  In order to obtain the kinked configuration we started with a configuration obtained from the kinked elastic rod, using a construction presented in \cite{HSS2011}.

\section{Conclusions}
\label{sec:6}

We have investigated minimum-energy configurations in the Skyrme-Faddeev and elastic rod models on $\RR^2\times S^1$ with Hopf degrees 1, 2 and 3.  We have found a good agreement between the two models, both qualitatively and often quantitatively.  For all charges and sufficiently large periods, the minimum-energy configuration in both models is a straight rod (or soliton).  Buckling occurs as the period is reduced.

For Hopf degree 1, the straight rod (or soliton) buckles slightly to form a helix in both models.  For Hopf degree 2, a much more visible buckling occurs and again the minimum energy configuration in both models at small periods is a helix.  Elastic rods with Hopf degree 3 undergo two successive buckling transitions, passing through a helix to form a kinked configuration at low periods.  The kinked configuration also appeared in simulations of the Skyrme-Faddeev model, although more numerical work is needed to determine whether or not it has a lower energy than a helix.

Our results show that the elastic rod model is a reliable description of Skyrme-Faddeev solitons.  They also demonstrate the utility of this model: without it, we would not have found the kinked configuration at Hopf degree 3.  Solitons in the Skyrme-Faddeev model are notoriously difficult to find numerically, particularly at high Hopf degree, and it is hoped that the elastic rod model will provide a useful tool for tackling this problem.

More generally, our results demonstrate a clear link between two models from field theory and elasticity theory.  They motivate the search for elastic phenomena in other field theories.

\bigskip
\noindent
\textbf{Acknowledgements} The work of DH was carried out at Durham University and supported by the Engineering and Physical Sciences Research Council (grant number EP/G038775/1).  A large part of the work of DF was carried out at the Dublin Institute for Advanced Studies (DIAS).  DF would like to thank D.H.\ Tchrakian for his help.

\bibliographystyle{utphys}
\bibliography{hopf}

\end{document}